\documentclass[twocolumn,
                showpacs,
                nofootinbib,
                nobibnotes,
                aps,
                superscriptaddress,
                amssymb,
                amsmath,
                floatfix,
                reprint,
                prx,
                notitlepage,
                showkeys,
                10pt,
		longbibliography]{revtex4-1}
                
\usepackage{amssymb}
\usepackage{amsmath}
\usepackage{cprotect}
\usepackage{graphicx}
\usepackage{dcolumn}
\usepackage{url}
\usepackage[colorlinks=true,breaklinks=true,allcolors=blue]{hyperref}
\usepackage{tikz}
\usepackage{dsfont}
\usepackage{epsfig}
\usepackage{color}
\usepackage[ruled]{algorithm2e}
\usepackage{bbold}
\usepackage[binary-units=true]{siunitx}
\usepackage{glossaries}
\usepackage{tikz}
\usepackage{bm}
\usepackage{hhline}
\usepackage{spverbatim}

\makeatletter
\let\cat@comma@active\@empty
\makeatother

\newcommand{\Fig}[1]{Fig.~\ref{fig:#1}}
\newcommand{\Tab}[1]{Tab.~\ref{tab:#1}}
\newcommand{\App}[1]{App.~\ref{app:#1}}

\begin{document}
\title{Miniaturizing neural networks for charge state autotuning in quantum dots}
\author{Stefanie Czischek}
\email{sczischek@uwaterloo.ca}
\affiliation{Department of Physics and Astronomy, University of Waterloo, Ontario, N2L 3G1, Canada}
\author{Victor Yon}
\affiliation{Institut Interdisciplinaire d’Innovation Technologique (3IT), Universit\'{e} de Sherbrooke, Sherbrooke J1K 0A5, Canada}
\affiliation{Laboratoire Nanotechnologies Nanosyst\`{e}mes (LN2) – CNRS UMI-3463 – 3IT, Sherbrooke J1K 0A5, Canada}
\affiliation{Institut quantique, Universit\'{e} de Sherbrooke, Sherbrooke J1K 2R1, Canada}
\author{Marc-Antoine Genest}
\affiliation{Institut quantique, Universit\'{e} de Sherbrooke, Sherbrooke J1K 2R1, Canada}
\affiliation{D\'{e}partement de Physique, Universit\'{e} de Sherbrooke, Sherbrooke J1K 2R1, Canada}
\author{Marc-Antoine Roux}
\affiliation{Institut quantique, Universit\'{e} de Sherbrooke, Sherbrooke J1K 2R1, Canada}
\affiliation{D\'{e}partement de Physique, Universit\'{e} de Sherbrooke, Sherbrooke J1K 2R1, Canada}
\author{Sophie Rochette}
\affiliation{Institut quantique, Universit\'{e} de Sherbrooke, Sherbrooke J1K 2R1, Canada}
\affiliation{D\'{e}partement de Physique, Universit\'{e} de Sherbrooke, Sherbrooke J1K 2R1, Canada}
\author{Julien Camirand Lemyre}
\affiliation{Institut quantique, Universit\'{e} de Sherbrooke, Sherbrooke J1K 2R1, Canada}
\affiliation{D\'{e}partement de Physique, Universit\'{e} de Sherbrooke, Sherbrooke J1K 2R1, Canada}
\author{Mathieu Moras}
\affiliation{Institut quantique, Universit\'{e} de Sherbrooke, Sherbrooke J1K 2R1, Canada}
\affiliation{D\'{e}partement de Physique, Universit\'{e} de Sherbrooke, Sherbrooke J1K 2R1, Canada}
\author{Michel Pioro-Ladri\`{e}re}
\affiliation{Institut quantique, Universit\'{e} de Sherbrooke, Sherbrooke J1K 2R1, Canada}
\affiliation{D\'{e}partement de Physique, Universit\'{e} de Sherbrooke, Sherbrooke J1K 2R1, Canada}
\affiliation{Institut Interdisciplinaire d’Innovation Technologique (3IT), Universit\'{e} de Sherbrooke, Sherbrooke J1K 0A5, Canada}
\affiliation{Laboratoire Nanotechnologies Nanosyst\`{e}mes (LN2) – CNRS UMI-3463 – 3IT, Sherbrooke J1K 0A5, Canada}
\author{Dominique Drouin}
\affiliation{Institut Interdisciplinaire d’Innovation Technologique (3IT), Universit\'{e} de Sherbrooke, Sherbrooke J1K 0A5, Canada}
\affiliation{Laboratoire Nanotechnologies Nanosyst\`{e}mes (LN2) – CNRS UMI-3463 – 3IT, Sherbrooke J1K 0A5, Canada}
\affiliation{Institut quantique, Universit\'{e} de Sherbrooke, Sherbrooke J1K 2R1, Canada}
\author{Yann Beilliard}
\affiliation{Institut Interdisciplinaire d’Innovation Technologique (3IT), Universit\'{e} de Sherbrooke, Sherbrooke J1K 0A5, Canada}
\affiliation{Laboratoire Nanotechnologies Nanosyst\`{e}mes (LN2) – CNRS UMI-3463 – 3IT, Sherbrooke J1K 0A5, Canada}
\affiliation{Institut quantique, Universit\'{e} de Sherbrooke, Sherbrooke J1K 2R1, Canada}
\author{Roger G. Melko}
\affiliation{Department of Physics and Astronomy, University of Waterloo, Ontario, N2L 3G1, Canada}
\affiliation{Perimeter Institute for Theoretical Physics, Waterloo, Ontario, N2L 2Y5, Canada}

\date{\today}

\begin{abstract}

A key challenge in scaling quantum computers is the calibration and control of multiple qubits.
In solid-state quantum dots, the gate voltages required to stabilize quantized charges are unique for each individual qubit, 
resulting in a high-dimensional control parameter space
that must be tuned automatically. Machine learning techniques are capable of processing high-dimensional data --
provided that an appropriate training set is available -- and have been successfully used for autotuning in the past.
In this paper, we develop extremely small feed-forward neural networks that can be used to detect 
charge-state transitions in quantum dot stability diagrams.  We demonstrate that these neural networks can be trained on 
synthetic data produced by computer simulations, and robustly transferred to the task of tuning an experimental device 
into a desired charge state. 
The neural networks required for this task are sufficiently small as to enable an implementation in
existing memristor crossbar arrays in the near future. This opens up the possibility of miniaturizing powerful
control elements on low-power hardware, a significant step towards on-chip autotuning in future quantum dot computers. 

\end{abstract}  

\maketitle

\section{Introduction}

Solid-state quantum dots (QDs) are one of several promising candidates for qubits, the basic building blocks of quantum computers 
\cite{Loss1998, Hanson2007, Veldhorst2014, Veldhorst2015, Maurand2016, Takeda2018, Yoneda2018, Watson2018}.
They are engineered in semiconductor devices by the electrostatic confinement of single charge carriers (electrons or holes),
and precisely tuned to a few-carrier regime where quantum effects dominate.
Even single-dot devices require the tuning of multiple gates for the control of reservoirs, dots, and tunnel barriers \cite{Hanson2007}. 
The relationship between applied gate voltages and physical properties of a QD is highly complex and device-specific, requiring significant calibration
and tuning.  Thus, a key challenge in scaling up QD architectures to act as multi-qubit devices will be tuning within the high-dimensional
space of gate voltages.  This is a highly non-trivial control problem that cannot be accomplished without significant automation \cite{Frees2019}. 

An automated process of finding a range of gate voltages in which a QD is in a specific carrier configuration is called autotuning.
Due to the variability inherent in different QD devices, autotuning naturally benefits from a data-driven approach.
Several compelling machine learning strategies have recently been introduced for from-scratch QD tuning \cite{Moon2020}, 
coarse tuning into charge regimes  \cite{Kalantre2019,Zwolak2020,Darulova2020,Darulova2020b,Durrer2020}, 
fine tuning couplings between multiple dots \cite{Teske2019,Esbroeck2020}, or performing autonomous measurements \cite{Lennon2019,Nguyen2020}.
These studies have demonstrated robust effectiveness in identifying electronic states and charge configurations,
and automating the precise tuning of gates. Common algorithms in supervised learning, such as support vector machines, deep, and convolutional
neural networks, are sufficiently powerful for the complex characterization tasks involved in this automation.
Coupled with the advent of community training data sets for tasks such as 
state recognition \cite{Zwolak2018} and charge transition identification \cite{Genest2020},
the enterprise of QD autotuning may well be the first demonstration of a large-scale qubit control problem tamed by machine learning.

Like many elements of modern microprocessor design, an important consideration in quantum computers will be 
integrating control technologies ``on-chip'' -- i.e.~on or near the physical qubits inside the cryostat.
This requires consideration of energy budgets to limit thermal dissipation in the various control tasks, the performance costs of transferring data, 
and the benefits of miniaturizing various control elements \cite{Vandersypen2017,Patra2018,Geck2019,Pauka2019}.
In autotuning of QDs, one of the simplest control tasks involves the identification of charge state transition lines in 
two-dimensional stability diagrams. 
While previous works use image analysis algorithms \cite{Lapointe2020} or deep (convolutional) neural networks \cite{Durrer2020,Kalantre2019}, 
in this paper we explore whether this task can be performed by extremely small feed-forward neural networks.
Mixed-signal vector-matrix multiplication accelerators based on crossbar arrays of emerging memory technologies
(e.g.~memristors \cite{Chua1971}) provide the possibility to implement 
sufficiently small neural networks on miniaturized hardware, which is able to reduce the power consumption by about three orders of magnitude compared to GPU implementations
\cite{Amirsoleimani2020,Sung2018,Alibart2013,Bayat2018,Hu2018,Sebastian2020,Zhang2020}.

Since these hardware implementations meet the requirements of an on-chip integration, we study the automated tuning of a single QD based on artificial 
neural networks which can be implemented on existing memristor crossbar arrays as a first step towards developing on-chip autotuning for general QD devices.
We find that neural networks with input layers as small as $5 \times 5$ pixels are capable of classifying small {\em patches} 
of a pre-processed stability diagram with sufficient accuracy to identify charge state transitions.
The patches can then be shifted around the stability diagram in order to
tune a quantum dot into a desired charge state starting from an arbitrary position.
One can increase the success rate of this autotuning procedure by considering arrays of connected patches.
By finding the minimal array size that provides high success rates, we show that the experimental measurement costs can be significantly reduced 
by covering only small regions of the stability diagram.
The number of parameters in the feed-forward neural networks required for this autotuning procedure are 
sufficiently small to make their implementation possible in present-day memristor crossbar arrays \cite{Sung2018,Sebastian2020,Amirsoleimani2020}.  
Further, we demonstrate that these parameters can be trained on synthetic (simulated) stability diagrams,
while achieving excellent performance in classifying transition lines from real experimental 
silicon metal-oxide-semiconductor quantum dot devices. Our work thus opens the possibility of taking advantage of the high speed and 
energy efficiency of memristor crossbar arrays \cite{Hu2018,Sebastian2020}, which could be integrated as part of a larger on-chip control system for 
QD autotuning in the near future.
While such memristor crossbar arrays can be used for the detection of charge transition lines, future work is required to enable the on-chip
integration of the full autotuning procedure, including tasks like signal processing and patch shifting.

\section{Supervised learning for stability diagrams}

In order to pursue a machine learning approach to the problem of quantum dot (QD) autotuning, we need to define training and testing data sets, 
and an appropriate neural network architecture.
Our end goal will be to use supervised learning to classify transition lines in small patches of the stability diagrams of
silicon metal-oxide-semiconductor QD devices.
Since experimental data for such stability diagrams is expensive to obtain, we aim to train our neural network 
architectures on synthetic data, which was obtained from a simulation package developed by M.-A.~Genest \cite{Genest2020}.
This synthetic data approximates experimental diagrams with noise effects after some signal processing, 
and provides ``ground truth'' labels for the charge stability sectors which can be used for training.

The machine learning architecture that we propose is a simple feed-forward neural network (FFNN), with a single output neuron
that indicates whether a transition line is found in the input patch of data or not.  In the following, we explore FFNNs 
with a very limited number of trainable parameters, roughly commensurate with the number of memristors
expected in high-density crossbar arrays available in the near future.  

\subsection{Data: experimental and synthetic}

\begin{figure*}
	\includegraphics[width=\textwidth]{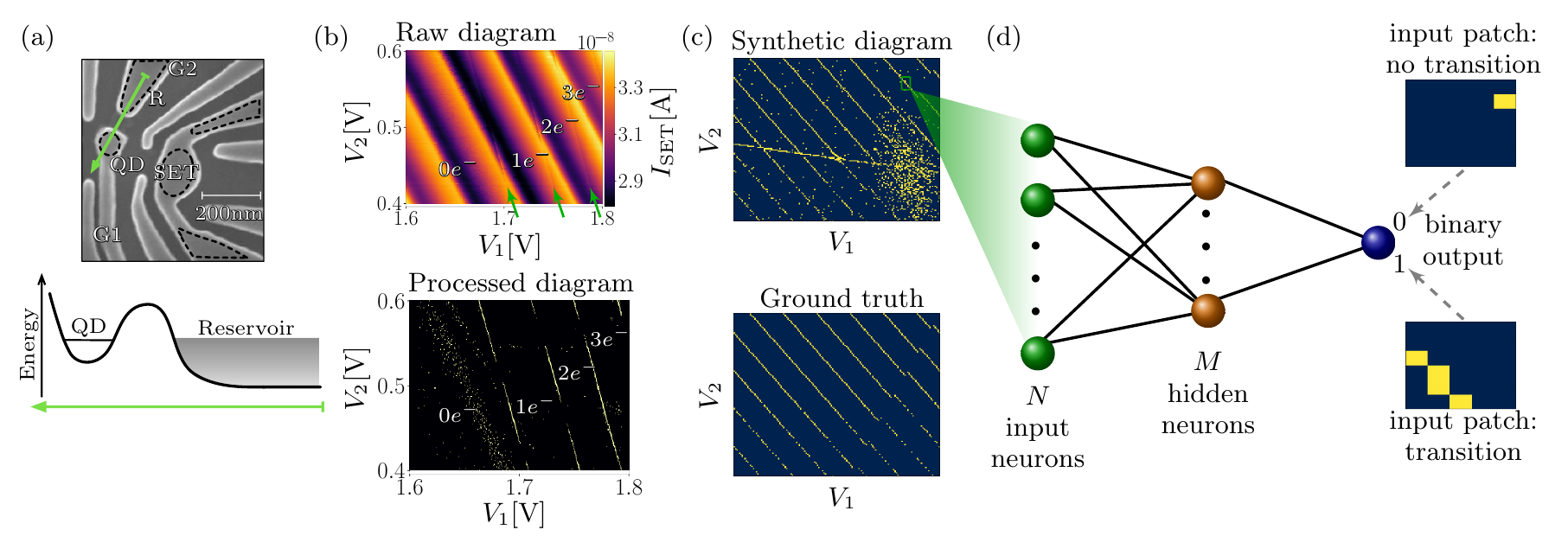}
	\caption{
	Steps towards autotuning experimental QD devices with artificial neural networks.
	(a) A scanning electron micrograph of the considered experimental single-QD device is shown in the upper panel. The QD is controlled via the electrostatic gate G1 and connected to an electron reservoir (R) whose electronic density is controlled via gate G2. 
	The SET is located next to the QD and enables the detection of transition lines. 
	Along the green arrow a potential landscape is created which is tuned to trap carriers in the QD.
	A sketch of the potential is shown in the lower panel, where the potential at the QD can be tuned via voltage $V_1$ applied at gate G1, and the potential at the reservoir can be tuned via voltage $V_2$ applied at gate G2, influencing the number of carriers trapped in the QD.
	(b) The experimentally measured stability diagram (upper panel) contains a strong oscillating background, where transition lines can be seen as sudden jumps which are indicated by green arrows as a guide to the eye. 
	By applying a signal processing algorithm, the background can be removed and a binary stability diagram shows the transition lines with additional noise effects (lower panel). 
	The transition lines separate different charge states, where the number of trapped electrons ($e^-$) in the QD is indicated for each regime.
	(c)  A numerical algorithm is used to create synthetic stability diagrams which simulate the experimental diagrams after the signal processing. 
	Realistic noise effects are applied to transition lines in the diagram (upper panel), while the ground truth data (lower panel) can be extracted containing the transition lines without noise. 
	(d) A small feed-forward neural network is trained on patches of the synthetically created stability diagrams to detect transition lines and is then applied on patches of experimental stability diagrams after signal processing. 
	The network input is given by the pixels in the small patch, where two example patches are plotted to the right of the network.
	The network output is a single binary neuron telling whether a transition line is detected in the input patch (output 1) or not (output 0). 
	We consider different numbers of hidden neurons in one or two layers in between input and output with the intention of miniaturizing the neural network and with this the computational complexity.}
	\label{fig:1}
\end{figure*}

Our experimental data is obtained from three 
silicon metal-oxide-semiconductor QD devices, with similar setups as previously discussed in \cite{Rochette2019}.
The upper panel of \Fig{1}(a) shows the corresponding setup of gate electrodes, creating a potential landscape along the green arrow. 
Two plunger gates at the QD and a connected reservoir (R) are denoted as G1 and G2.
The resulting potential landscape for the carriers (electrons or holes) is sketched in the lower panel of \Fig{1}(a), where the low-potential island corresponds to the QD in which carriers are trapped.
We consider a unipolar QD device, which can only trap either electrons or holes. 
While all following statements are as well true for holes, we focus on considering electrons throughout the paper.
The number of electrons trapped in the single QD defines its charge state, which can be tuned most efficiently via the gate G1 controlling the depth of the QD potential.
Neighboring gates, such as G2 which defines the electron reservoir connected to the QD, can additionally affect the QD charge state through cross-capacitance effects \cite{Rochette2019}.

Transitions between different charge states can be measured via a single-electron transistor (SET) which is tuned such that a measured current $I_{\mathrm{SET}}$ is sensitive to potential changes.
Changing the charge state of the QD by adding or removing an electron causes abrupt jumps in $I_{\mathrm{SET}}$.
The current $I_{\mathrm{SET}}$ can be measured as a function of the voltages at the QD plunger gate ($V_1$ applied at gate G1) and the electron reservoir gate ($V_2$ applied at gate G2), while keeping all other gate voltages fixed, providing two-dimensional charge {\em stability diagrams} \cite{Hanson2007}.
Transitions between different charge states appear as distinct lines in the stability diagram and must be identified in order to tune the QD \cite{Rochette2019, Lapointe2020, Durrer2020, Kalantre2019}.

The upper panel of \Fig{1}(b) shows a measured stability diagram where an additional oscillating background, 
caused by cross-capacitance effects of G1 and G2 acting on the SET, impedes detection of the transition lines.
This background is specific for the considered measurement setup \cite{Rochette2019} and caused by the absence of a compensation procedure such as dynamic feedback control \cite{Yang2011}.
It can be removed via a signal processing algorithm as discussed in \cite{Lapointe2020}, transforming the stability diagram into a binary image, see lower panel of \Fig{1}(b).
To obtain a binary stability diagram with clear charge transition lines, a high-pass filter is first used to remove fluctuating background effects from the raw experimental signal, where the cutoff frequency can be extracted from a Fourier transform.
In a second step, a Hilbert transform is applied to extract the frequency of the observed oscillations, where jumps indicating charge transition lines appear as negative peaks in the obtained frequency.
By choosing a suitable threshold, the transition lines can directly be found in the processed signal, leading to binary stability diagrams with bright pixels indicating charge transitions.
While we apply the signal processing algorithm to full pre-measured stability diagrams in the following, it can as well be applied to smaller voltage patches.

After the signal processing, noise effects still appear in the transformed diagrams.
This noise is partly due to imperfections in the experimental setup, such as the SET coupling to external charges which are not part of the quantum dot, and partly due to the signal processing algorithm, 
where for example negative peaks in the frequency after the Hilbert transform are not detectable at the extrema of the oscillating background \cite{Rochette2019, Lapointe2020}.
It is these noisy transition lines, which lie in the stability diagrams after signal processing, that we aim to detect with the FFNN
of the next section.
Detecting the transition lines in the binary image is a general approach and can be applied to any experimental setup after a setup-specific signal 
processing algorithm has been applied to the raw measurement outcome.

It would be preferable to have a large database of labelled experimental stability diagrams with which to train supervised machine learning
strategies such as the considered FFNN.  Since a suitable database is not available, however, we propose to train the networks on synthetic data.
We use the numerical algorithms discussed in \cite{Genest2020} to simulate post-signal-processed
stability diagrams, which include noise effects similar to those found in experimental measurements, see \App{1} for details.
This enables us to create a large data set of synthetic stability diagrams, which include the clear presence of transition lines [see \Fig{1}(c)].
This can then be transformed into a training set for supervised learning on small patches of pixels,
which include the ground-truth labels corresponding to the presence or absence of a transition line in the patch.  
To discuss this further, we must first examine in more detail the specific neural network architecture used for patch classification.

\subsection{Neural network training strategy}

Motivated by on-chip integration of the autotuning inside the cryostat, we wish to explore the performance of supervised learning tasks on the minimal-size
artificial neural networks possible.
Therefore, in this section we restrict ourselves to small FFNNs with one or two hidden layers between the input and the output layer, as illustrated in \Fig{1}(d).
We furthermore restrict the amount of hidden neurons to small numbers, which puts limitations on the complexity of the network.
In order to know which small network architectures are useful, we will explore their performance for 
a simple classification task on the smallest possible patches of the binary stability diagram after signal processing.

The input neurons of the network correspond to the pixels in the patch, and a single output neuron is used for a binary classification of whether a transition line is detected or not.
The output neuron takes the value one if a transition line is detected in the input patch and is zero otherwise.
All neurons of neighboring layers are connected, and each connection is weighted with an individual parameter.
These connecting weights, and additional bias parameters for each hidden neuron, form the set of learnable parameters that are adapted when training the FFNN.
We apply sigmoid activation functions to all neurons, such that the output corresponds to the probability of detecting a transition line.
In order to get a binary classification, we round the output to zero or one.

During the training procedure, patches of synthetic stability diagrams are shown to the FFNN and the binary classification output is compared to the
 known true classification.
The network parameters are then updated by propagating the classification error backwards, where we use the Adam optimizer \cite{Kingma2014}
 on a cross-entropy loss function with a learning rate $\eta=0.001$ to calculate the parameter updates.
In each training epoch, all elements of the synthetic training data set are shown to the network and the 
parameters are updated accordingly.
We apply the parameter updates resulting from batches of $50$ input patches at a time.
Once the network is trained, the learned parameters are fixed and we test the network performance
 by classifying a test data set of patches extracted from pre-processed experimentally measured stability diagrams and comparing the outcome 
 to manually labelled data.

As experimental stability diagrams contain large areas without transition lines [see \Fig{1}(b)], 
when creating a test data set, the overhead of empty patches compared to patches with transition lines should be considered.
To account for this, we add a weighting towards patches with transition lines in the training procedure to compensate.
When evaluating the performance of the FFNN classifier in the next section,
we consider the total accuracy on the full test set, as well as the accuracies for correctly classifying only patches with and without transition lines.

Finally, by shifting patches across the diagram via an algorithm driven by the classification outcome, the quantum dot can be tuned into any desired charge state \cite{Lapointe2020, Durrer2020}.
As only charge transition lines can be detected but not the absolute charge value, this shifting algorithm needs to find the regime where the quantum dot is empty, which is reached when no more transition lines can be found [$0e^-$ in \Fig{1}(b)].
From this reference point the QD can be filled with the desired number of electrons by crossing the corresponding number of transition lines.
We discuss this algorithm as a step towards on-chip autotuning in Sec.~\ref{TuningResults}.

The entire autotuning process then consists of three essential steps. First an experimentally measured signal over a small regime of gate voltages is processed to create a
binary input for the FFNN. 
The pre-trained network then classifies this input to determine whether a transition line has been detected, which is indicated by a binary network output. 
Depending on the classification outcome, the considered gate voltage regime is shifted according to a pre-defined shifting algorithm in order to find the single-electron regime. 
These steps are repeated until the desired charge state of the QD device is found. 
While we focus on using miniaturized FFNNs for the classification task to provide a first step towards an on-chip integration of QD autotuning, the signal processing and shifting
algorithm steps still require computations performed on a classical computer, outside the cryostat. 
Thus, future work is required to develop cryogenic electronics and in-situ signal processing entabling a complete integration of the autotuning procedure as close as possible to the QD device.

\section{Results}

In this section, we begin by discussing the detection of transition lines in small patches of the stability diagram using small FFNNs.
In the below, we train the network on a data set of $80000$ patches which are extracted from $800$ numerically created synthetic stability diagrams at random positions.
The trained network is then tested on $2700$ patches extracted from random positions of $27$ experimentally measured stability diagrams after signal processing which are labeled manually, see \Fig{6} in \App{1}.
Due to the thin transition lines, the stability diagrams mainly consist of areas without lines and thus only about $7\%$ of the training patches and $1.5\%$ of the test patches contain transition lines, depending on the patch size.
This significant overhead of empty patches requires the weighting towards patches with transition lines when training the FFNN.
The experimental stability diagrams are measured on three different QD devices with similar setups.
The neural network classifications are then used to define a shifting algorithm for small patches, capable of autotuning a single QD into a desired charge state when starting from a random position in the stability diagram.

\subsection{Detection of transition lines}

\begin{figure}
	\includegraphics[width=\linewidth]{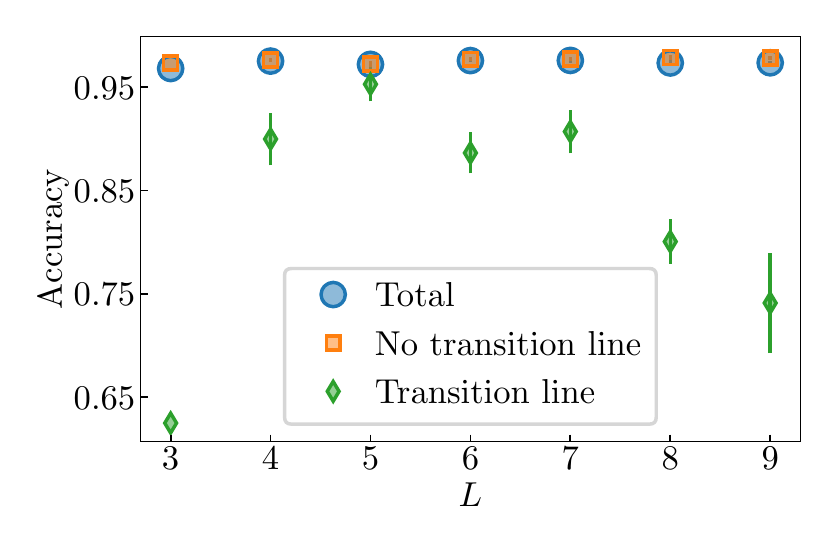}
	\caption{
	Classification accuracy as a function of the patch size $N = L \times L$ pixels.
	The total test accuracy (blue) is considered together with the accuracy for correctly classifying empty patches (orange) or patches including a transition line (green).
	Five feed-forward neural networks with a single hidden layer of $10$ neurons are trained over $500$ epochs for each data point.
	Accuracies are averaged over training epochs $450$ to $500$ to ensure convergence for larger patch sizes (see \Fig{7} in \App{1}).
	The results of all five networks are averaged with error bars denoting standard deviations.
	}
	\label{fig:2}
\end{figure}

To begin our supervised learning procedure, we find a suitable patch size by analyzing the accuracy reached 
by a network consisting of a single hidden layer with ten hidden neurons.
Generally the patch size depends on the resolution of the measured stability diagram, which defines the voltage step corresponding to one pixel.
The stability diagrams considered in this work all have similar resolutions of about $5\times10^{-4}$V per pixel for $V_1$ and $5\times10^{-3}$V per pixel for $V_2$.
For convenience we stick to the expression of pixels, as we associate each input neuron in the FFNN with one pixel of the considered patch.

\Fig{2} shows the accuracies reached when classifying patches of $N = L\times L$ pixels with varying $L$.
The total accuracy and the accuracy for correctly classifying empty patches always reach $\sim96\%$.
In contrast to this, the accuracy for correctly detecting transition lines is lower and shows a strong dependence on the patch size.
We find the best performance for $L=5$, where all three accuracies are high.
Therefore, we focus on this patch size, which dictates the size of the input layer of the FFNNs in the following.
For larger patches the classification accuracy decreases since we keep the number of hidden neurons in the FFNN fixed and small.
Correctly classifying larger patches requires larger and more complex network structures.
On the other side, the classification accuracy decreases for smaller patch sizes, where it is harder to detect structures.
For a correct classification, the patch needs to be large enough to clearly capture the transition line structure so that it can be distinguished from unstructured noise pixels.

\begin{figure}
	\includegraphics[width=\linewidth]{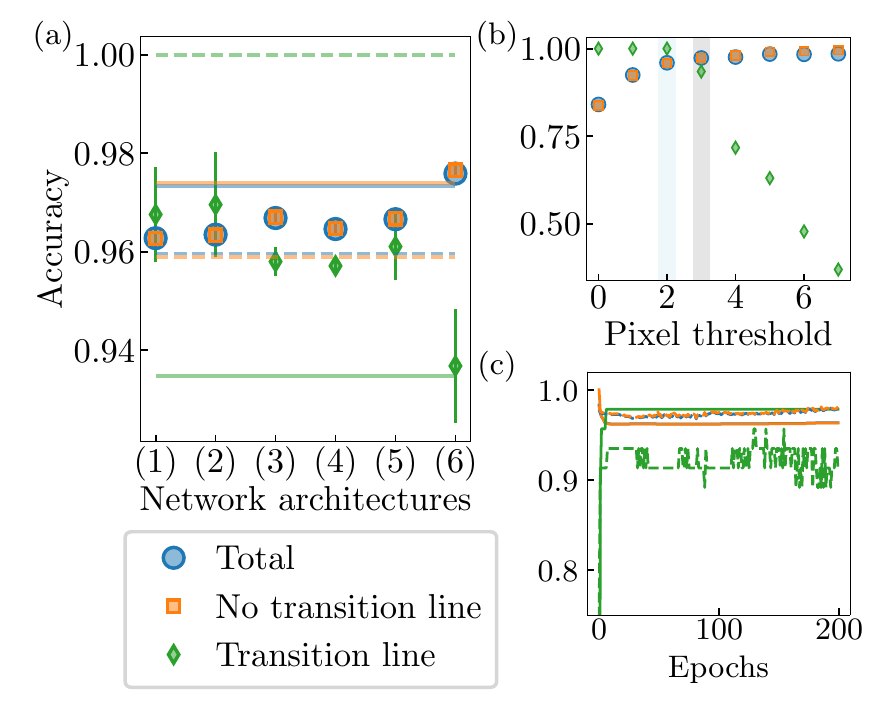}
	\caption{
	Classification accuracy with different network setups.
	In all panels the total accuracy (blue), as well as the accuracy for correctly classifying patches with (green) and without (orange) transition lines are shown.
	All $y$-axes show the accuracy, where labels are omitted for clarity in panels (b) and (c). 
	(a) Accuracy as a function of the network architecture as stated in \Tab{1} for a patch of size $L=5$.
	For each data point five feed-forward neural networks are trained for $200$ epochs and the accuracies over the last $50$ epochs are averaged, with standard deviations shown by the error bars.
	The accuracies are compared to results of a pixel classifier with a threshold of $2$ pixels [dashed lines, blue area in (b)] and a threshold of $3$ pixels [solid lines, gray area in (b)].
	(b) Accuracy of the pixel classifier on the experimental stability diagrams after signal processing as a function of the threshold pixel number. 
	High values for all accuracies are reached for a threshold of $2$ pixels (blue area), as well as for a threshold of $3$ pixels (gray area). 
	(c) Classification accuracy as a function of training epochs in a single run using network setup (2) (solid lines) and (6) (dashed lines), and a patch of size $L=5$. 
	Convergence is found after less than $100$ training epochs, justifying a training over $200$ epochs for $L=5$.}
	\label{fig:3}
\end{figure}

\begin{table}
	\begin{tabular}{cccc}
	\hhline{====}
	Network & \# hidden & \# hidden & \# learnable\\
	architecture & layers & neurons & parameters\\
	\hline
	(1) & 1 & $M_1=5$ & $135$\\ 
	(2) & 1 & $M_1=10$ & $270$ \\
	(3) & 1 & $M_1=15$ & $405$ \\
	(4) & 2 & $M_1=5$, $M_2=10$ & $200$ \\
	(5) & 2 & $M_1=10$, $M_2=5$ & $320$ \\
	(6) & 2 & $M_1=100$, $M_2=100$ & $12800$\\
	\hhline{====}
	\end{tabular}
	\caption{Feed-forward neural network architectures considered in this manuscript. 
	The number of input neurons in all networks is given by the number of pixels in the considered patch and is $L\times L=25$ for $L=5$.
	Each network has a single output neuron telling whether a transition line is found in the patch or not. 
	In between the input and the output layer we consider a single [(1)-(3)] or two [(4)-(6)] layers of $M_1$ hidden neurons in the first and $M_2$ hidden neurons in the second layer.
	The number of learnable parameters corresponds to $25\times M_1+M_1$ connecting weights and $M_1$ biases for setups (1)-(3), 
	and $25\times M_1+M_1\times M_2+M_2$ connecting weights and $M_1+M_2$ biases for setups (4)-(6).}
	\label{tab:1}
\end{table}

With the size of the input and output layers thus fixed, the total number of learnable parameters in the FFNN is dictated by the number of hidden units.
In \Fig{3}(a) we analyze the dependence of the classification accuracy on different hidden unit architectures.
We consider different network setups as summarized in \Tab{1}, consisting of a single [(1)-(3)] or two [(4)-(6)] hidden layers of different sizes.
In addition, to ensure that the small number of hidden neurons does not have a significant limiting effect on the expressivity of the network,
we compare these results to a network with two hidden layers of $100$ neurons each [(6)], presumed to be representative of large neuron numbers.
We train all network setups for $200$ epochs on the synthetic training set.

As illustrated in \Fig{3}(a),
the total accuracy and the accuracy for correctly classifying empty patches reaches $\sim96\%$ for all architectures in \Tab{1}.
However, a clear dependence on the network architecture is observed in the accuracy for classifying patches with transition lines.
Interestingly, optimal results are found for networks with small numbers of learnable parameters [(1) and (2)], affirming that we are not limited by the restricted number of hidden neurons.
The accuracy decreases for larger network setups with more learnable parameters.

In addition, we compare the accuracies reached by the neural network classifier with results from a
simple classifier based on the amount of bright pixels in the patch.
If the amount of bright pixels crosses a certain threshold defined via a pixel number, the patch is classified as containing a transition line.
\Fig{3}(b) shows the accuracies reached with this pixel classifier as a function of the threshold pixel number.
Good performances are found when choosing the threshold at $2$ or $3$ pixels.
In \Fig{3}(a) we add the results of the pixel classifier for a direct comparison and observe that small networks, especially architectures (1) and (2), outperform the pixel classifier.
This further justifies the use of FFNNs to learn to detect structure in the small patches.

Finally, to confirm that training the networks for $200$ epochs is sufficient, \Fig{3}(c) shows the accuracies achieved
as function of the number of training epochs, for network architectures (2) and (6) trained on a $5\times 5$ pixel patch.
We find convergence after less than $100$ training epochs, providing evidence to justify our choice.

To conclude this section, we refer again to \Fig{3}(a) to emphasize that optimal performance is consistent with network architecture (2).
As illustrated there, using this FFNN on a $5\times 5$ pixel input patch, we reach $\sim96\%$ for all testing accuracies.
Given the fact that this high test accuracy for classifying experimental data occurs for a small and simple network structure, 
particularly one that is trained on simulated synthetic data, we argue that the results are quite promising.
Therefore, we emphasize this architecture in the next section, 
to define a shifting algorithm for small patches, with the end goal of autotuning a QD into a desired charge state in the stability diagram.

\subsection{Tuning the device with a patch shifting algorithm} \label{TuningResults}
In this section we consider a shifting algorithm which we develop for the small patches considered above.
With this algorithm we tune a single QD into a desired charge state when starting at a random position in the stability diagram.
The full algorithm is discussed in detail in \App{2} and depends on the classification outcome of the FFNN for the patch at each position.
While we apply the algorithm on finite pre-measured stability diagrams, we develop it in a way such that it can be generally applied 
to infinite stability diagrams, enabling an in-situ application on experimental measurements after signal processing.
The algorithm hence works for stability diagrams of arbitrary sizes, but due to the definition in terms of pixels it strongly depends on the relation between pixels and gate voltage steps.
The diagrams considered in this work are obtained from three similar devices and show similar gate voltage steps per pixel.
However, a general application of the shifting algorithm requires a transformation between the pixel and voltage values for the considered QD device.
As the transition lines only provide information about changes in the charge state but not the absolute charge of the QD, we first need to find 
a reference point in the diagram \cite{Rochette2019, Kalantre2019, Lapointe2020, Durrer2020}.
Similar to \cite{Lapointe2020, Durrer2020} we use the empty QD as a reference point, where no electrons are trapped in the QD.
This point is reached when no more transition lines are detected while shifting the patch to the left.

From this reference point, the QD can be filled with the desired number of electrons by crossing the corresponding number of transition lines,
as identified by the neural network strategy discussed in the previous section.
Below, we focus on finding the single-electron regime, where the QD can be interpreted as a qubit.

\begin{figure}
	\includegraphics[width=\linewidth]{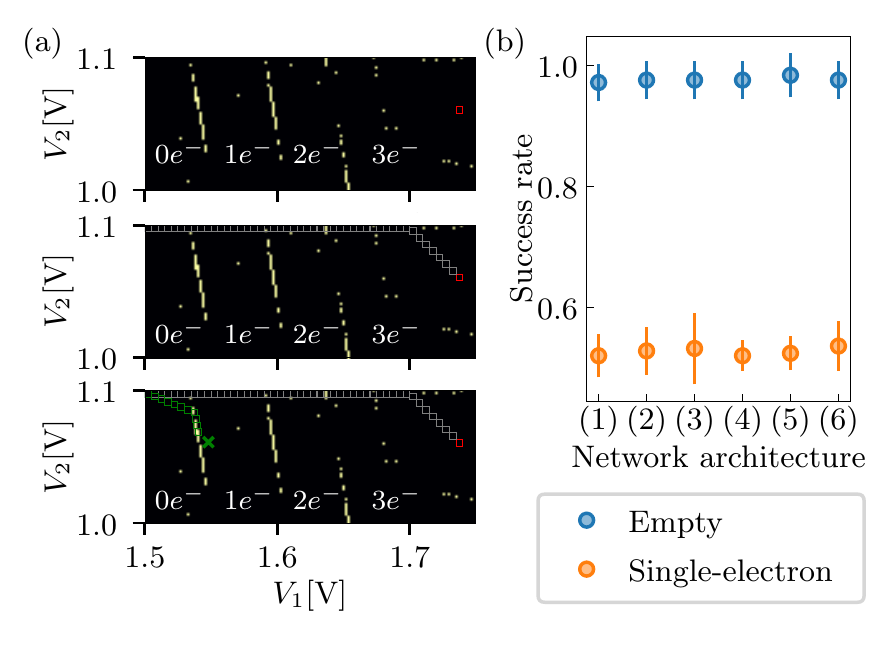}
	\caption{
	Shifting the patch into the single-electron regime.
	(a) Sketch of the shifting algorithm with a $5\times 5$ pixel patch classified with a feed-forward neural network with setup (2) (see \Tab{1}). 
	The upper panel shows the randomly chosen start point as a red patch. 
	In the center panel the gray patches show the path towards the reference point where the QD is empty, which is found in the upper left corner. 
	The full path is shown in the lower panel with green patches indicating the shifting towards the single-electron regime. 
	The green cross denotes the final position. 
	The different charge states separated by the transition lines are indicated with the corresponding number of electrons ($e^-$).
	(b) Success rate for finding the empty (blue) and the single-electron regime (orange) as a function of the network architectures as listed in \Tab{1}. 
	For each point $25$ experimentally measured stability diagrams are considered after signal processing and the success rate is averaged over ten randomly chosen initial points per diagram with the error bars denoting standard deviations. 
	The start points are random but the same for all network architectures. 
	The success rates are high for finding the reference points where the QD is empty, but low for finding the single-electron regimes independent of the network architectures.}
	\label{fig:4}
\end{figure}
An example for the shifting algorithm finding the single-electron regime is shown in \Fig{4}(a), where a patch of $5\times 5$ pixels is classified with an FFNN with architecture (2) (see \Tab{1}).
A patch is placed at a randomly chosen initial position (red patch, upper panel) and shifted towards the empty regime (gray patches, center panel).
To find the reference point where the QD is empty, the patch is first shifted diagonally to the upper left in order to find the regime where transition lines are present.

Once the first transition line is detected, the vertical position is fixed and the patch is shifted to the left until the next transition line is found.
When the patch has been shifted for a certain amount of steps without detecting a transition line, a reference point where the QD is empty is found, 
since no more electrons can be removed.

Here we apply the shifting algorithm on pre-measured finite stability diagrams and the movement of the patch is limited by the diagram borders.
While in \Fig{4}(a) the reference point is hence found on the upper left corner, these border effects do not appear for an in-situ autotuning, where the voltages can be tuned further.
Due to these limitations by the diagram borders, we can not ensure that the QD is empty in the found reference point.
Thus, we interpret the regime with minimum electron filling of the QD in the finite stability diagrams as empty QD regime.
This issue does not appear in the ideal case of infinite stability diagrams, where the patch can be shifted further and we can ensure that no more transition lines are detected
when finding the empty QD regime.

From the found reference point where the QD is empty, the patch is shifted to the right until the first transition line is detected to fill the dot with a single electron (green patches, lower panel).
The patch follows the line for a few steps to avoid misclassifications of noise effects.
The target point in the single-electron regime is indicated by a green cross in the lower panel of \Fig{4}(a) and is chosen to the right of the detected transition line.

To analyze the performance of the shifting algorithm we consider $25$ experimentally measured stability diagrams after signal processing where the true single-electron regime can be clearly identified.
For each diagram we choose ten random initial positions and check how successful the shifting algorithm performs in finding the reference point and the single-electron regime.

We perform this analysis on all the different FFNN architectures listed in \Tab{1}.
\Fig{4}(b) shows the success rates averaged over the ten initial positions.
Error bars denote the standard deviation when averaging over the initial positions, and show that the dependence on the latter is small.
The reference point is found successfully in $\sim98\%$ of the diagrams, while the success rate for finding the single-electron regime is rather small with $\sim53\%$.
Finding the empty QD regime is easier as in most cases the left corner of the diagram is reached, where the algorithm terminates.
This is a specific effect of considering finite pre-measured stability diagrams and is not expected to appear when directly tuning the experimental device according to the shifting algorithm.
The fact that no clear dependence on the network architecture can be observed suggests that the performance of the shifting algorithm is the limiting factor.

The high classification accuracies suggest that the simple control task of identification of charge state transition lines can be parlayed into 
a successful algorithm for tuning to a specific charge sector of the QD stability diagram, while using very small neural networks
suitable e.g.~for implementation on vector-matrix multiplication accelerators.
However, an obvious caveat of choosing the small patch size can be seen in the diagrams in \Fig{4}(a) and \Fig{1}(b).
The transition lines are interrupted by gaps caused by the signal processing procedure which shows difficulties in detecting transition lines at the extrema of the oscillating background, see \App{1} for details.
These gaps are larger than the classified patch, so that transition lines can be missed.
While striving to keep the input of the FFNN small, this issue can be overcome, for example, by extending the algorithm to 
couple adjacent patches.
To illustrate this, we consider an array of $K\times K$ patches, which reduces the risk of the patch translation algorithm
hitting a gap in the transition lines, while only slightly increasing the computational costs and preserving the high accuracy reached in the classification of small patches.
Each patch in the array is classified individually by our small FFNN, and the shifting algorithm in this case
now depends on whether a transition line is found in any patch of the array.
Here it is sufficient to detect a transition line in at least one individual patch to classify the entire array as containing a transition line for the shifting algorithm, see \App{2} for details.

\begin{figure}
	\includegraphics[width=\linewidth]{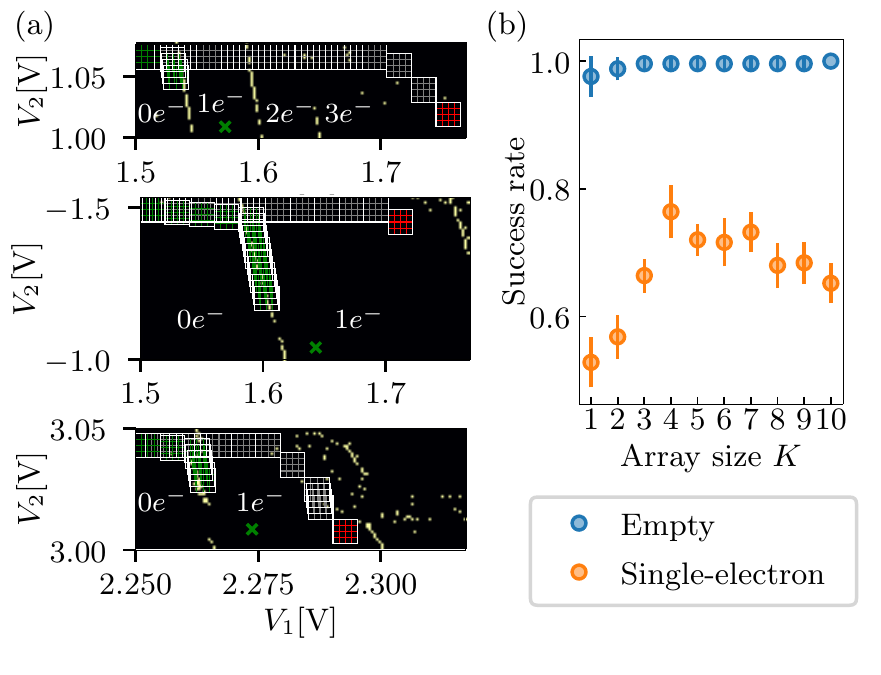}
	\caption{
	Shifting patch arrays to find the single-electron regime.
	(a) Shifting algorithm for an array of $4\times 4$ patches of $5\times 5$ pixels each on three different stability diagrams. 
	Start points are shown in red, the shifting path towards the reference point of an empty QD is shown in gray, and green denotes the shifting path towards the single-electron regime with the green cross indicating the final position. 
	A jump to the lower right is performed after detecting the transition line from the reference point to ensure ending up in the single-electron regime. 
	The number of electrons ($e^-$) is indicated for the charge states separated by the transition lines.
	(b) Success rate for finding the empty (blue) and the single-electron regime (orange) using an array of $5\times 5$ pixel patches which are classified with a feed-forward neural network with structure (2) (see \Tab{1}) as a function of array size $K$.
	Arrays are quadratic arrangements of patches with $K$ stating the number of patches in one direction. 
	We consider $25$ experimentally measured stability diagrams after signal processing and average the success rates over ten start points for each diagram with error bars denoting standard deviations. 
	Initial points are chosen randomly but are the same for all array sizes. 
	Empty QD regimes are always found with good accuracy while the success rates for the single-electron regime show a strong dependence on the array size.}
	\label{fig:5}
\end{figure}
\Fig{5}(a) illustrates the shifting algorithm for an array of $4\times 4$ patches with $5\times 5$ pixels per patch on three different stability diagrams.
The whole array is shifted to first find the reference point where the QD is empty and then detect the first transition line, leading to a target point in the single-electron regime indicated by the green cross.

\Fig{5}(b) shows the success rate of the shifting algorithm as a function of the array size $K$, where the same data as in \Fig{4}(b) is used.
An FFNN with architecture (2) is used for the classification of the $5\times 5$ pixel patches.
The reference point where the QD is empty is found successfully in $\sim99\%$ of the diagrams, while finding the single-electron regime shows a dependence on the array size.
An increase in the success rate can be observed already when going from $K=1$ (single patch) to $K=2$ and a further increase is found for larger $K$.
At $K=4$ the success rate reaches its maximum and saturates before it appears to decrease again for $K\geq 8$.
The tuning into the single-electron regime is successful in $\sim75\%$ of the diagrams for $K=4$, which is a relatively large success rate.
For different patch sizes $L$ with similarly high classification accuracies (see \Fig{2}), we find similar success rates if we adapt the array size $K$ such that the total amount 
of $(L\times K)^2$ pixels in the array is kept constant.
We hope this result will encourage the further exploration of small FFNNs to autotune general QD devices.

\section{Discussion and outlook}
We have demonstrated that the task of autotuning a single quantum dot (QD) into a target charge state 
can be robustly achieved by harnessing the power of very small feed-forward neural networks (FFNNs).
Such FFNNs are the workhorses of supervised machine learning, and enable a data-driven approach 
akin to transfer learning, where networks trained on simulated data can be used for classification of experimental QD data.
We have shown in particular that a classification approach, where such small neural networks are trained to detect the
presence or absence of charge transition lines in small patches of stability diagrams, can be transferred with excellent accuracy
to pre-processed experimental stability diagrams obtained from silicon metal-oxide-semiconductor quantum dot devices.
Further, by combining this classification approach with an algorithm that shifts arrays of small input patches,
we have shown that a single QD can successfully be autotuned into the single-electron regime, when starting from a
random configuration of gate voltages.
With our small FFNNs, we reached classification accuracies, as well as final success rates, which are comparable to previous results with deep (convolutional) neural networks \cite{Kalantre2019, Durrer2020}.
The success of small input patches in combination with the possibility to pre-process smaller areas of stability diagrams further suggests 
that the experimental cost of measuring the stability diagram can be significantly reduced.
Indeed, with our shifting algorithm we are able to consider smaller regions than in comparable works \cite{Lapointe2020,Durrer2020}. 

We have observed excellent success rates for finding the single-electron regime with our shifting algorithm for arrays of small patches.
This algorithm has been designed in a way such that we expect it to be generally applicable to infinitely large stability diagrams, suggesting an in-situ tuning of experimental devices.
We have demonstrated its success on finite sized pre-measured data, where the shifting algorithm is further limited by the borders of the stability diagrams.
These borders make it impossible for us to ensure that the QD is empty, so that we define the single electron regime as the regime with the smallest number of 
electrons in the finite data, which is trivially found at the left border of the stability diagram.
We expect smaller success rates for in-situ autotuning since infinite voltage ranges make it harder to find the empty QD regime and one might not be able to
find the region of clear transition lines from any random initial voltage setting.
However, infinite diagrams enable the possibility to ensure that the QD is empty at the reference point by tuning over a larger regime of gate voltages.
The actual application of the proposed shifting algorithm on infinite stability diagrams remains an open topic for future work.

We have found that the FFNNs used for the classification task require a very small number of learnable parameters 
to achieve a maximal classification accuracy.  This suggests that such neural networks could 
be implemented on state-of-the-art memristor crossbar arrays \cite{Sung2018,Sebastian2020,Amirsoleimani2020}.
Hence, our work provides a first step towards developing an energy-efficient autotuning device, which could
conceivably be implemented in an on-chip control system for QDs.
In order to further pursue this idea, the performance of real memristor crossbar arrays, which is expected to be influenced by imperfections \cite{Adam2018,Wang2019}, 
will need to be carefully analyzed on the classification task at hand.

While we focussed on autotuning a device with a single QD, the application of quantum computing gates requires devices with two or more QDs.
For these systems the tuning procedure becomes more complex and is even more in need of an automatic algorithm.
Thus, while our work provides a first step towards developing an on-chip autotuning device, future work requires the study of the suggested algorithm
on systems of multiple QDs.

It is particularly important to emphasize that throughout this work we have trained the small FFNNs on input patches 
extracted from synthetic stability diagrams generated with a numerical simulation package \cite{Genest2020}.
Due to the relative expense involved in obtaining experimental data on QDs, the success of this transfer learning approach 
is an important step in further developing our machine learning strategy.  It also exposes the clear opportunity
for creating a larger community data set for training neural networks to detect charge transition lines in QDs, similar to \cite{Zwolak2018}.
Eventually, such data sets -- whether synthetic or experimental -- will play an important role in standardizing and
benchmarking machine-learning based QD control and autotuning.

Finally, we remark that in this work we applied our charge transition line detection algorithm on stability diagrams that have already 
undergone significant signal processing according to \cite{Lapointe2020}.  With our successful machine learning approach, 
this signal processing step becomes computationally more expensive than the classification task.
It would therefore be interesting to further refine our data-driven strategy, to 
eventually enable a similar neural-network based classification to occur directly on raw experimental data.
It is feasible that significant miniaturization of such tasks could lead to a highly-efficient on-chip control 
system for autotuning quantum dots in the near future.

\section*{Acknowledgments}

We thank Z. Bandic for many enlightening discussions.
We thank our collaborators at Sandia National Laboratories for providing the samples used for the development of our algorithm.
RGM is supported by the Canada Research Chair (CRC) program 
and the Perimeter Institute for Theoretical Physics. Research at Perimeter Institute is supported in part by the Government of Canada through the 
Department of Innovation, Science and Economic Development Canada and by the Province of Ontario through the Ministry of Colleges and Universities. 
This research was undertaken thanks in part to funding from the Canada First Research Excellence Fund.
The work was supported in part by the Natural Sciences and Engineering Research Council of Canada (NSERC), the NSERC Alliance program, PROMPT Qu\'ebec and Fonds de recherche Nature et technologies of Qu\'ebec.

\section*{Data availability and implementation}
The code to create synthetic stability diagrams is obtainable at \href{https://github.com/Marc-AntoineGenest/Quantum-dots-simulator-and-image-processing-toolbox}{GitHub}\cprotect\footnote{\spverb|https://github.com/Marc-AntoineGenest/Quantum-dots- simulator-and-image-processing-toolbox|},
and the data set used for training is available at \href{https://github.com/quantumdata/QD-charge-classification}{GitHub}\cprotect\footnote{\spverb|https://github.com/quantumdata/QD-charge-classification|}.
The remaining data that support the findings of this study are available from the correspomding author upon reasonable request.
The feed-forward neural networks are implemented and trained using PyTorch \cite{Paszke2019} and NumPy \cite{Harris2020}, figures are created using Matplotlib \cite{Hunter2007}.

\setcounter{equation}{0}
\setcounter{figure}{0}
\setcounter{table}{0}
\setcounter{page}{1}
\makeatletter
\renewcommand{\theequation}{A\arabic{equation}}
\renewcommand{\thefigure}{A\arabic{figure}}
\renewcommand{\theHequation}{\theequation}
\renewcommand{\theHfigure}{\thefigure}
\appendix
\section{Data creation} 
\label{app:1}
The training of the feed-forward neural network to detect transition lines in patches of stability diagrams requires a large training data set.
As it is hard to create such a large set of experimentally measured stability diagrams, we create the training set numerically.
However, we obtain the reached classification accuracies from applying the trained network to an experimentally measured test data set.

To get stability diagrams from the experimental device, the quantum dot is coupled to a single-electron transistor (SET).
A current $I_{\mathrm{SET}}$ running through the SET is measured while tuning the voltages at two gate electrodes to obtain two-dimensional diagrams.
The remaining gate electrodes are kept at fixed voltages, as two-dimensional diagrams bring many advantages for the autotuning procedure compared to one-dimensional single-gate scans \cite{Lapointe2020}.

The current $I_{\mathrm{SET}}$ shows an oscillatory behavior where transition lines are obtained as jumps in the oscillations \cite{Rochette2019, Lapointe2020}.
Since the transition lines are hard to extract from this raw data, we separate them from the oscillatory current via a signal processing algorithm as discussed in \cite{Lapointe2020}.
In this algorithm, the raw signal is first sent through a high-pass filter to remove background effects.
Afterwards the frequency of the oscillations is extracted via a Hilbert transform.
At charge state transitions, which appear as jumps in the oscillations, the frequency shows negative peaks which can be identified by a threshold method.
The considered threshold is adapted to the obtained frequency distribution.
This algorithm provides a binary mapping of the stability diagrams, where the transition lines are extracted from the raw data.

While those binary diagrams already contain the extracted charge transition lines which are needed to tune the device into a desired charge state, the diagrams also contain noisy pixels due to imperfections in the experimental measurement procedure and in the signal processing algorithm.
The main imperfections that appear are the following:
\begin{itemize}
\item The SET couples to an external charge: an additional line appears in the diagram which is not a transition line,
\item the quantum dot couples to an external charge: the transition lines experience a sudden jump and continue at a shifted position,
\item measurements are performed too fast: if the tunneling rate is too low, the electrons do not get through the barrier during the measurement, leading to a spreading and fading of transition lines at low voltages,
\item when the derivative of the oscillating signal is close to zero, which is the case at the top and bottom of the oscillations, the negative peaks in the frequency do not appear: quasi-periodic gaps are found in the transition lines,
\item the measurement sensitivity can decrease: spots of noisy pixels appear.
\end{itemize}
All these effects make the detection of transition lines in the binary diagrams harder, which is why we use the feed-forward neural network to detect them.

The network performance is tested on experimental diagrams after signal processing, but we train it on numerically created synthetic diagrams which simulate the noise effects discussed above.
The synthetic training data can be generated efficiently and faster than experimental data and additionally comes with ground truth data containing only the transition lines without noise.
This ground truth data is necessary to perform the supervised training of the network, where labels need to be provided.

To create the synthetic training data, we use the algorithm discussed in \cite{Genest2020}.
First, ideal noiseless diagrams are created which contain the transition lines simulating a given experimental setup.
The positions of the transitions are calculated via the Thomas-Fermi approximation and hence show a realistic orientation and spacing \cite{Genest2020, Kalantre2019, Zwolak2020}.
This simulation directly provides binary diagrams which are similar to the experimental diagrams after signal processing.
The noise effects discussed above are then added numerically to the ideal data, so that synthetic diagrams close to the experimental ones are created.
Additionally, we add a unitary background noise by turning each pixel bright with probability $0.01$.

Keeping track of the transition line transformations when applying the noise effects yields the ground truth data.
We take into account that some noise effects, such as gaps appearing in the lines or distortions of lines, also apply to the ground truth data, while other effects, such as additional lines, noisy spots, or the background noise, do not affect the ground truth.

With this algorithm we create $20$ ideal stability diagrams with the parameters of the experimental setup chosen randomly in a specific regime.
We then apply $20$ randomly chosen combinations of the discussed noise effects to each ideal diagram, leading to $400$ synthetic stability diagrams.
As the concentration of transition lines is high in the synthetic data, we additionally create $400$ stability diagrams where we randomly apply the noise effects of the SET coupling to an external charge or the measurement losing its sensitivity to empty diagrams without lines and add the background noise.
This yields a total of $800$ synthetic stability diagrams from which patches can be extracted to train the feed-forward neural network on.

\begin{figure}
	\includegraphics[width=\linewidth]{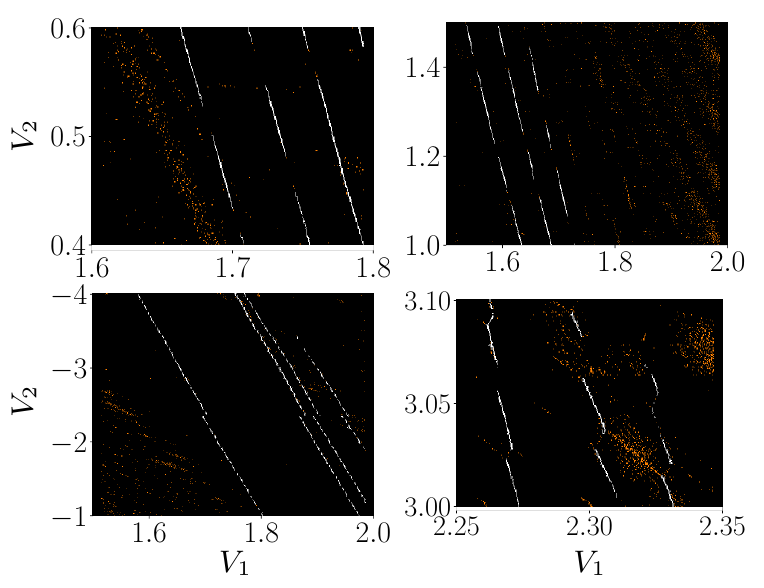}
	\caption{Examples for manually labeled stability diagrams. 
	The four panels show experimentally measured stability diagrams after the signal processing algorithm.
	Bright pixels have manually been labeled as transition lines (white pixels) and noise (orange pixels), while dark pixels denote the background.
	The network classification accuracy is determined by comparing the output of the trained neural network with the manually labelled true data.
	}
	\label{fig:6}
\end{figure}

Once the neural network is trained, we test its accuracy on processed experimental stability diagrams.
In order to determine the classification performance, we manually label these diagrams and distinguish between transition lines and noise pixels, 
see \Fig{6} for examples of labelled stability diagrams.
This data allows us to compare the network classification of small patches with the true labels to extract the classification accuracy on experimental data.

\begin{figure}
	\includegraphics[width=\linewidth]{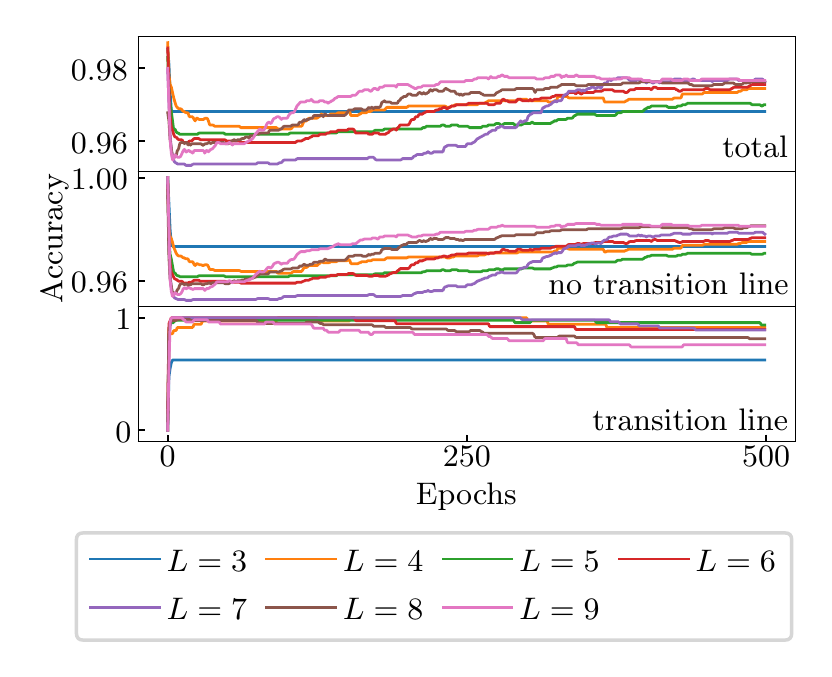}
	\caption{Neural network classification accuracy as a function of training epochs for different input patch sizes.
	The total accuracy (upper panel) and the accuracies for correctly classifying patches without (centre panel) and with (lower panel) transition lines are plotted
	as a function of the network training epochs for different sizes of input patches with $L\times L$ pixels (different colors).
	Convergence is observed for $\geq 400$ training epochs in all cases.
	}
	\label{fig:7}
\end{figure}

\Fig{7} further shows the classification accuracy of the trained feed-forward neural network as a function of training epochs for different patch sizes consisting of $L\times L$ pixels.
The three panels show the total accuracy (upper panel), as well as the accuracy for correctly classifying patches without (center panel) and with (lower panel) transition lines.
While larger patch sizes require more training epochs to converge, we observe that all networks converge to stable classification accuracies for more than $400$ epochs.
This justifies the choice of considering accuracies after $450$--$500$ training epochs in \Fig{2} in the main text for all patch sizes.

\section{Shifting algorithm}
\label{app:2}
The quantum dot (QD) device is tuned into the single-electron regime via shifting the classified small patch over the stability diagram.
The algorithm for the shifting procedure depends on the classification outcome of the considered patch and can be divided into two parts.
First, the reference point where the QD is empty needs to be found, from which the device can be tuned into the single-electron regime in a second step.
This algorithm is similar to \cite{Durrer2020,Lapointe2020}, but needs to be adapted for the small patch sizes.

In the following we generally talk about an array of $K\times K$ patches as considered in the main text, where the special case of a single patch corresponds to $K=1$.
Each patch has a size of $L\times L$ pixels.
We develop the shifting algorithm based on pixels in the stability diagram, which for the devices considered in the main text corresponds to steps of $5\times 10^{-4}$V per 
pixel for the voltage $V_{1}$ applied at gate $G_1$ (horizontal in the stability diagrams), and $5\times 10^{-3}$V per pixel for the voltage $V_{2}$ applied at gate $G_2$ 
(vertical in the stability diagrams), see \Fig{1}(a) in the main text.
The shifting algorithm is hence specific for these QD devices and requires a translation between pixel values and voltage steps when generally applied to different device setups.

To find the empty QD regime we start with an array at a random position in the stability diagram and search for the region where transition lines are clearly present.
We then diagonally shift the array $K\times L$ pixels to the left and the same amount of pixels upwards as long as no transition line is detected.
While the upper corner of the diagram is not reached, we apply periodic boundary conditions in the horizontal direction when searching for the first transition line.
With this we ensure to find the region of clear transition lines in the finite pre-measured stability diagrams, while a better guess for an initial position would be necessary in the ideal
case of infinite stability diagrams for in-situ autotuning.

If a line is detected in at least one patch of the array, we have found the transition line region and we follow the line by shifting the array one pixel to the left and $L$ pixels upwards.
If the upper boundary of the diagram is reached, we shift the patch $L$ pixels to the left and lose the line.
To avoid ending up in a wrong regime due to misclassifications of noise, we only declare the transition line as found if the same patch detects the line ten steps in a row.
Otherwise, if the line is not detected anymore, we interpret the observation as noise and continue searching for the first transition line via diagonal shifts.

When the first transition line has been confirmed, we follow it until it fades out.
We then shift the array $K\times L$ pixels to the left until any patch detects the next line, which we follow again.
When no line is detected after moving $40/K$ steps of $K\times L$ pixels to the left, no electrons are removed anymore and the dot is empty.
We define this position as the reference point in the empty QD regime.

During the whole procedure of finding the empty-dot regime, the process is terminated if the upper left corner of the diagram is reached and this corner is defined as being in the empty regime.
All shifts are only applied as long as the boundaries of the stability diagram are not crossed, otherwise the array is not shifted in this direction.

Once having found the empty QD regime, we fill the QD with a single electron, which is done by crossing the left-most transition line.
Since the reference point is to the left of all transition lines, the array is shifted $K\times L$ pixels to the right and two pixels down as long as no line is detected.

If at least one patch in the array detects a transition line, we follow the line and shift the array one pixel to the right and $L$ pixels downwards.
When the bottom boundary of the diagram is reached, we continue shifting the patch $L$ pixels to the right.
We apply the same procedure as before to avoid conclusions drawn from misclassifications and only declare a line as detected when a single patch finds it five times in a row.
Otherwise we continue shifting the array until the next line is detected.
Since the shifting towards the reference point moves the patch to the upper part of the diagram, the chances of hitting the bottom boundary before finding a transition line are very low.

When the first transition line is found starting from the empty regime, we move the array diagonally $2K\times L$ pixels to the right and downwards and with this find a position in the single-electron regime.
If a different charge state is desired, the procedure can be iterated analogously until the desired number of transition lines is crossed.
The shifting process for filling the dot is terminated whenever a corner of the diagram is reached and the position at this corner is defined as the single-electron regime.

Mind that the terminations due to reaching corners which we apply in both steps of the shifting algorithm are specific for the case of having finite pre-measured stability diagrams.
Generally, for in-situ tuning of experimental devices, the voltage limitations are not expected to be reached.


%

\end{document}